\documentclass[preprint,showpacs,preprintnumbers,amsmath,amssymb]{revtex4}
\usepackage{amssymb}
\usepackage{amsmath}
\usepackage{graphicx}
\usepackage{epsfig}
\usepackage{subfigure}
\usepackage{amsfonts}
\usepackage{CJK}

\begin{document}

\title{Universal quantum gate with hybrid qubits in circuit quantum electrodynamics}

\author{Chui-Ping Yang$^{1}$}
\email{yangcp@hznu.edu.cn}
\author{Zhen-Fei Zheng$^{2}$}
\author{Yu Zhang$^{3}$}
\address{$^1$Department of Physics, Hangzhou Normal University, Hangzhou 310036, China}
\address{$^2$Key Laboratory of Quantum Information, University of Science and Technology of China, Heifei 230026, China}
\address{$^3$School of Physics, Nanjing University, Nanjing 210093, China}


\begin{abstract}
Hybrid qubits have recently drawn intensive attention in quantum computing. We here propose a method to implement a universal controlled-phase gate of two hybrid qubits via two three-dimensional (3D) microwave cavities coupled to a superconducting flux qutrit. For the gate considered here, the control qubit is a microwave photonic qubit (particle-like qubit), whose two logic states are encoded by the vacuum state and the single-photon state of a cavity, while the target qubit is a cat-state qubit (wave-like qubit), whose two logic states are encoded by the two orthogonal cat states of the other cavity. During the gate operation, the qutrit remains in the ground state; therefore decoherence from the qutrit is greatly suppressed. The gate realization is quite simple, because only a single basic operation is employed and neither classical pulse nor measurement is used. Our numerical simulations demonstrate that with current circuit QED technology, this gate can be realized with a high fidelity. The generality of this proposal allows to implement the proposed gate in a wide range of physical systems, such as two 1D or 3D microwave or optical cavities coupled to a natural or artificial three-level atom. Finally, this proposal can be applied to create a novel entangled state between a particle-like photonic qubit and a wave-like cat-state qubit.
\end{abstract}
\maketitle
\date{\today }

Quantum gates, acting on hybrid qubits (i.e., different types of qubits),
have attracted tremendous attention, because of their importance in
connecting quantum information processors with different encoding qubits as
well as their significant application in transferring quantum states between
a quantum processor and a quantum memory. In recent years, many theoretical
proposals have been presented for realizing a universal two-qubit
controlled-phase (CP) or controlled-not (CNOT) gate with various hybrid
qubits, such as a cat-state qubit and a charge qubit [1], a flying photonic
qubit and an atomic qubit [2], a charge qubit and an atomic qubit [3], a
spin qubit and a Majorana qubit [4], a photonic qubit and a superconducting
qubit [5], and so on. Moreover, the two-qubit CP or CNOT gate with a flying
optical photon and a single trapped atom [6], as well as the two-qubit CP
gate with a $^{40}$Ca$^{+}$ qubit and one $^{43}$Ca$^{+}$ qubit [7] have
been demonstrated in experiments.

Circuit QED, consisting of microwave cavities and superconducting (SC)
qubits, has been considered as one of the most promising candidates for
quantum computing [8,9]. Besides SC qubits, microwave photonic qubits
(encoded in the photon-number states) and cat-state qubits (encoded in
superposition of coherent states) are two types of important qubits for
quantum information processing (QIP) and quantum communication.
Particularly, cat-state qubits have drawn intensive attention due to their
enhanced life times [10]. Recently, much progress has been made for quantum
state engineering and QIP with microwave photonic qubits [11-14] or
cat-state qubits [15-18].

The goal of this letter focuses on realizing a two-qubit CP gate with two
hybrid qubits, i.e., a microwave photonic qubit and a cat-state qubit, based
on a circuit-QED [Fig.1(a)]. The two-qubit CP gate considered here is
described by
\begin{align}
|0\rangle |cat\rangle & \rightarrow |0\rangle |cat\rangle ,\text{ }|0\rangle
|\overline{cat}\rangle \rightarrow |0\rangle |\overline{cat}\rangle ,
\nonumber \\
\text{ }|1\rangle |cat\rangle & \rightarrow |1\rangle |cat\rangle ,\text{ }%
|1\rangle |\overline{cat}\rangle \rightarrow -|1\rangle |\overline{cat}%
\rangle ,
\end{align}%
where $|cat\rangle $ and $|\overline{cat}\rangle $ are two orthogonal cat
states, representing the two logic states of a cat-state qubit, while $%
|0\rangle $ and $|1\rangle $ are the two logic states of a microwave
photonic qubit. Eq. (1) implies that when the control qubit (first
qubit) is in the state $|1\rangle ,$ a phase flip happens to the state $|%
\overline{cat}\rangle $ of the target qubit (second qubit). It is known that
a two-qubit CP gate, together with single-qubit gates, forms a set of
universal gates for quantum computing [19].

This proposal has the following advantages. During the gate operation, the
coupler qutrit remains in the ground state and thus decoherence from the
qutrit is greatly suppressed. The gate realization is quite simple because
only one-step operation is needed. Moreover, neither classical pulse nor
measurement is required. Our numerical simulation shows that high-fidelity
implementation of the gate is feasible with the current circuit QED
technology. This proposal can be extended to realize the proposed gate with
two 1D or 3D microwave or optical cavities coupled to a natural or
artificial three-level atom.

Note that a two-qubit CP gate can be easily transferred to a two-qubit CNOT
gate, by performing a Hadamard gate on the target qubit before and after the
two-qubit CP gate [19]. Therefore, our proposal can also be applied to
realize a hybrid two-qubit CNOT gate, described by $|0\rangle |cat\rangle
\rightarrow |0\rangle |cat\rangle ,$ $|0\rangle |\overline{cat}\rangle
\rightarrow |0\rangle |\overline{cat}\rangle ,$ $|1\rangle |cat\rangle
\rightarrow |1\rangle |\overline{cat}\rangle ,$ and $|1\rangle |\overline{cat%
}\rangle \rightarrow |1\rangle |cat\rangle ,$ acting on the two hybrid
qubits. To the best of our knowledge, how to realize a two-qubit CP or CNOT
gate with a microwave photonic qubit and a cat-state qubit has not been
reported yet.

The two-qubit CP or CNOT gate here allows to create a novel entangled state $%
\left\vert 0\right\rangle \left\vert cat\right\rangle +\left\vert
1\right\rangle \left\vert \overline{cat}\right\rangle $, through first
preparing a microwave photonic qubit in the state $\left\vert 0\right\rangle
+\left\vert 1\right\rangle $ while a cat-state qubit in the state $%
\left\vert cat\right\rangle $ and then applying the above-mentioned
two-qubit CNOT gate. This type of entangled state provides the first test of
a Bell inequality violation between a particle-like photonic qubit and a
wave-like cat-state qubit and has applications in hybrid quantum communication.
Recently, hybrid entanglement $\left\vert
0\right\rangle \left\vert \alpha \right\rangle +\left\vert 1\right\rangle
\left\vert -\alpha \right\rangle $ between a particle-like photonic qubit
(encoded with $\left\vert 0\right\rangle $ and $\left\vert 1\right\rangle $)
and a wave-like coherent-state qubit (encoded with the coherent states $%
\left\vert \alpha \right\rangle $ and $\left\vert -\alpha \right\rangle $)
or between quantum and classical states of light [20,21] has been
demonstrated in experiments, which has drawn increasing attention because
hybrid entanglement of light is a key resource in establishing hybrid
quantum networks and connecting quantum processors with different encoding
qubits.

\begin{figure}[tbp]
\begin{center}
\includegraphics[bb=33 398 540 582, width=8.5 cm,height=3.5cm, clip]{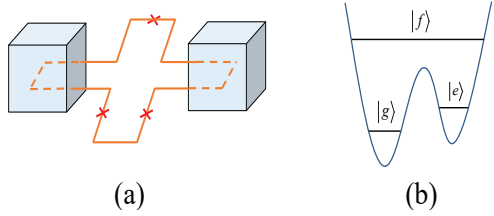} \vspace*{%
-0.28in}
\end{center}
\caption{(Color online) (a) Diagram of two 3D cavities inductively coupled
to a superconducting flux qutrit. The qutrit consists of three Josephson
junctions and a superconducting loop. (b) Level configuration of the flux
qutrit, for which the transition between the two lowest levels can be made
weak by increasing the barrier between two potential wells.}
\label{fig:1}
\end{figure}

Consider two 3D microwave cavities inductively coupled to a SC flux qutrit
[Fig.~1(a)]. The qutrit has three levels $|g\rangle $, $|e\rangle $ and $%
|f\rangle $ [Fig.~1(b)]. The $|g\rangle \leftrightarrow $ $|e\rangle $
transition is weak due to the barrier between the two potential wells. Cavity $1$ is dispersively coupled to the $|g\rangle
\leftrightarrow |f\rangle $ transition with coupling constant $g_{1}$ and
detuning $\delta _{1}$ but highly detuned (decoupled) from the $|e\rangle
\leftrightarrow |f\rangle $ transition of the qutrit. In addition, cavity $2$
is dispersively coupled to the $|e\rangle \leftrightarrow |f\rangle $
transition with coupling constant $g_{2}$ and detuning $\delta _{2}$ but
highly detuned (decoupled) from the $|g\rangle \leftrightarrow |f\rangle $
transition of the qutrit (Fig.~2). These conditions can be met by prior
adjustment of the qutrit's level spacings or the cavity frequency. For a SC
qutrit, the level spacings can be rapidly (within 1-3 ns) tuned [22]. The
frequency of a microwave cavity can be rapidly adjusted with a few
nanoseconds [23].

\begin{figure}[tbp]
\begin{center}
\includegraphics[bb=181 479 417 736, width=4.5 cm,height=4.3cm,clip]{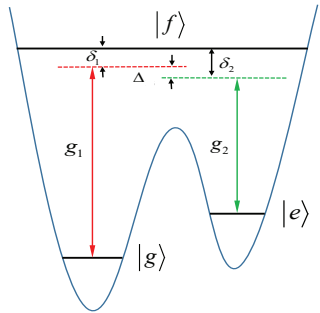}
\vspace*{-0.20in}
\end{center}
\caption{(Color online) Cavity $1$ is dispersively coupled to the $|g\rangle
\leftrightarrow |f\rangle $ transition of the qutrit with coupling strength $%
g_{1}$ and detuning $\protect\delta _{1}$, while cavity $2$ is dispersively
coupled to the $|e\rangle \leftrightarrow |f\rangle $ transition of the
qutrit with coupling strength $g_{2}$ and detuning $\protect\delta _{2}$.
The red vertical line represents the frequency $\protect\omega _{c_{1}}$ of
cavity $1,$ while the blue vertical line represents the frequency $\protect%
\omega _{c_{2}}$ of cavity $2$.}
\label{fig:2}
\end{figure}

Under the above assumptions, the Hamiltonian of the whole system in the
interaction picture and after the rotating-wave approximation, is given by
(in units of $\hbar =1$)
\begin{equation}
H_{\mathrm{I}}=g_{1}(e^{-i\delta _{1}t}\hat{a}_{1}^{+}\sigma
_{fg}^{-}+h.c.)+g_{2}(e^{-i\delta _{2}t}\hat{a}_{2}^{+}\sigma
_{fe}^{-}+h.c.),
\end{equation}%
where $\hat{a}_{1}$ ($\hat{a}_{2}$) is the photon annihilation operator of
cavity $1$ ($2$)$,$ $\sigma _{fg}^{-}=|g\rangle \langle f|$, $\sigma
_{fe}^{-}=|e\rangle \langle f|$, $\delta _{1}=\omega _{fg}-\omega _{c_{1}}>0,
$ and $\delta _{2}=\omega _{fe}-\omega _{c_{2}}>0$ (Fig. 2). Here, $\omega
_{c_{1}}$ ($\omega _{c_{2}}$) is the frequency of cavity $1$ ($2$); while $%
\omega _{fg},$ $\omega _{fe},$ and $\omega _{eg}$ are the $|f\rangle
\leftrightarrow |g\rangle ,$ $|f\rangle \leftrightarrow |e\rangle ,$ and $%
|e\rangle \leftrightarrow |g\rangle $ transition frequencies of the qutrit,
respectively.

By applying the large-detuning conditions $\delta _{1}\gg g_{1}$ and $\delta
_{2}\gg g_{2}$, the Hamiltonian (2) can be written as [24]
\begin{align}
H_{\mathrm{e}}=& -\lambda _{1}\hat{n}_{1}|g\rangle \langle g|-\lambda _{2}%
\hat{n}_{2}|e\rangle \langle e|+\left( \lambda _{1}+\lambda _{2}+\lambda _{1}%
\hat{n}_{1}+\lambda _{2}\hat{n}_{2}\right) |f\rangle \langle f|  \nonumber \\
& -\lambda (e^{i\bigtriangleup t}\hat{a}_{1}^{+}\hat{a}_{2}\sigma
_{eg}^{-}+h.c.),
\end{align}%
where $\sigma _{eg}^{-}=|g\rangle \langle e|,$ $\lambda
_{1}=g_{1}^{2}/\delta _{1}$, $\lambda _{2}=g_{2}^{2}/\delta _{2}$, $\lambda
=\left( g_{1}g_{2}/2\right) (1/\delta _{1}+1/\delta _{2})$, $\bigtriangleup
=\delta _{2}-\delta _{1}=\omega _{c_{1}}-\omega _{c_{2}}-\omega _{eg}$ (Fig.
2); $\hat{n}_{1}=\hat{a}_{1}^{+}\hat{a}_{1}$ and $\hat{n}_{2}=\hat{a}_{2}^{+}%
\hat{a}_{2}$ are the photon number operators for cavities $1$ and $2$,
respectively. The terms in the last line of Eq. (3) describe the $|e\rangle $
$\leftrightarrow $ $|g\rangle $ coupling induced by the two-cavity
cooperation. For $\bigtriangleup \gg \{\lambda _{1},\lambda _{2},\lambda \}$%
, the Hamiltonian (3) becomes [24]
\begin{align}
H_{\mathrm{e}}& =-\lambda _{1}\hat{n}_{1}|g\rangle \langle g|-\lambda _{2}%
\hat{n}_{2}|e\rangle \langle e|+\left( \lambda _{1}+\lambda _{2}+\lambda _{1}%
\hat{n}_{1}+\lambda _{2}\hat{n}_{2}\right) |f\rangle \langle f|  \nonumber \\
& -\chi \hat{n}_{1}\left( 1+\hat{n}_{2}\right) |g\rangle \langle g|+\chi
\left( 1+\hat{n}_{1}\right) \hat{n}_{2}|e\rangle \langle e|
\end{align}%
where $\chi =\lambda ^{2}/\Delta $. When the levels $|e\rangle $ and $%
|f\rangle $ are initially not occupied, they will remain unpopulated because
the Hamiltonian (4) does not induce both $\left\vert g\right\rangle
\rightarrow \left\vert e\right\rangle $ and $\left\vert g\right\rangle
\rightarrow \left\vert f\right\rangle $ transitions. Hence, this Hamiltonian
(4) reduces to
\begin{equation}
H_{\mathrm{e}}=-\lambda _{1}\hat{n}_{1}|g\rangle \langle g|-\chi \hat{n}%
_{1}\left( 1+\hat{n}_{2}\right) |g\rangle \langle g|.
\end{equation}

Assume that the qutrit is initially in the ground state $\left\vert
g\right\rangle $. It will remain in this state because the Hamiltonian (5)
cannot induce any transition for the qutrit. Therefore, the Hamiltonian (5)
reduces to
\begin{equation}
\widetilde{H}_{\mathrm{e}}=-\eta \hat{n}_{1}-\chi \hat{n}_{1}\hat{n}_{2},
\end{equation}%
where $\eta =\lambda _{1}+\chi .$ Here, $\widetilde{H}_{\mathrm{e}}$ is the
effective Hamiltonian governing the dynamics of the two cavities. The
unitary operator $U=e^{-i\widetilde{H}_{\mathrm{e}}t}$ is expressed as
\begin{equation}
U=\exp \left[ i\eta \hat{n}_{1}t\right] \otimes \exp \left( i\chi \hat{n}_{1}%
\hat{n}_{2}t\right) .
\end{equation}

Let us now consider two hybrid qubits $1$ and $2$. Qubit $1$ is a microwave
photonic qubit (particle-like qubit), whose two logic states are represented
by the vacuum state $|0\rangle $ and the single-photon state $|1\rangle $ of
cavity $1$. Qubit $2$ is a cat-state qubit (wave-like qubit), whose two
logic states are represented by the two orthogonal cat states $|cat\rangle
=M_{\alpha }^{+}(|\alpha \rangle +|-\alpha \rangle )$ and $|\overline{cat}%
\rangle =M_{\alpha }^{-}(|\alpha \rangle -|-\alpha \rangle )$. Here, $%
M_{\alpha }^{\pm }=1/\sqrt{2(1\pm e^{-2|\alpha |^{2}})}$ are normalization
coefficients. In terms of $|\alpha \rangle =e^{-|\alpha
|^{2}/2}\sum\limits_{n=0}^{\infty }\frac{\alpha ^{n}}{\sqrt{n!}}|n\rangle $
and $|-\alpha \rangle =e^{-|\alpha |^{2}/2}\sum\limits_{n=0}^{\infty }\frac{%
(-\alpha )^{n}}{\sqrt{n!}}|n\rangle $, we have
\begin{equation}
|cat\rangle =\sum\limits_{m=0}^{\infty }C_{2m}|2m\rangle ,\ \ |\overline{cat}%
\rangle =\sum\limits_{n=0}^{\infty }C_{2n+1}|2n+1\rangle ,
\end{equation}%
where $C_{2m}=2M_{\alpha }^{+}e^{-|\alpha |^{2}/2}\alpha ^{2m}/\sqrt{(2m)!}$
and $C_{2n+1}=2M_{\alpha }^{-}e^{-|\alpha |^{2}/2}\alpha ^{2n+1}/\sqrt{%
(2n+1)!}$. Here, $m$ and $n$ are non-negative integers. From Eq.~(8), one can see that the state $|cat\rangle $ is
orthogonal to the state $|\overline{cat}\rangle $, which is independent of $\alpha $
(except for $\alpha =0$).

Based on Eq. (7) and Eq. (8), one can easily see that the unitary operator $%
U $ leads to the following state transformation

\begin{align}
& |0\rangle _{1}|cat\rangle _{2}\overset{U}{\longrightarrow }|0\rangle
_{1}|cat\rangle _{2}  \nonumber \\
& |0\rangle _{1}|\overline{cat}\rangle _{2}\overset{U}{\longrightarrow }%
|0\rangle _{1}|\overline{cat}\rangle _{2}  \nonumber \\
& |1\rangle _{1}|cat\rangle _{2}\overset{U}{\longrightarrow }%
\sum\limits_{m=0}^{\infty }\exp [i\eta t]\exp [i2m\chi t]C_{2m}|1\rangle
_{1}|2m\rangle _{2},  \nonumber \\
& |1\rangle _{1}|\overline{cat}\rangle _{2}\overset{U}{\longrightarrow }%
\sum\limits_{n=0}^{\infty }\exp [i\eta t]\exp [i\left( 2n+1\right) \chi
t]C_{2n+1}|1\rangle _{1}|2n+1\rangle _{2},
\end{align}%
where subscripts $1$ and $2$ represents qubits $1$ and $2$. For $\chi t=\pi $ and $\eta t=2k\pi $ ($k$ is a positive integer), Eq.~(9)
becomes
\begin{eqnarray}
&&|0\rangle _{1}|cat\rangle _{2}\overset{U}{\longrightarrow }|0\rangle
_{1}|cat\rangle _{2}  \nonumber \\
&&|0\rangle _{1}|\overline{cat}\rangle _{2}\overset{U}{\longrightarrow }%
|0\rangle _{1}|\overline{cat}\rangle _{2}  \nonumber \\
&&|1\rangle _{1}|cat\rangle _{2}\overset{U}{\longrightarrow }|1\rangle
_{1}|cat\rangle _{2},\text{ }  \nonumber \\
&&|1\rangle _{1}|\overline{cat}\rangle _{2}\overset{U}{\longrightarrow }%
-|1\rangle _{1}|\overline{cat}\rangle _{2},
\end{eqnarray}%
which shows that when the control qubit $1$ is in the state $\left\vert
1\right\rangle $, a phase flip (from sign $+$ to $-$) happens to the state $|%
\overline{cat}\rangle $ of the target qubit $2$. The state transformation
(10) shows that a two-qubit CP gate, described by Eq. (1), is implemented by
the above operation.

For the two-qubit controlled phase gate described in Eq. (1) or Eq. (10), the control qubit and the target qubit can exchange their functions. Namely, when the control qubit is a cat-state qubit and the target qubit is a microwave photonic qubit, the two-qubit controlled phase gate can still be realized with the above operation.

From the above description, it can be seen that the coupler qutrit remains
in the ground state $\left\vert g\right\rangle $ during the entire
operation. Hence, decoherence from the qutrit is greatly suppressed. In
addition, the gate is realized with a single basic operation described by
the unitary operator $U.$

In above, we have set $\chi t=\pi $ and $\eta t=2k\pi ,$ resulting in
\begin{equation}
g_{2}=\frac{2\delta _{2}}{\delta _{1}+\delta _{2}}\sqrt{\frac{\delta
_{1}\Delta }{2k-1}}.
\end{equation}%
In practice, the coupling strength $g_{2}$ can be adjusted by a prior design
of the sample with appropriate capacitance or inductance between the qutrit
and cavity $2$.

\begin{figure}[tbp]
\begin{center}
\includegraphics[bb=184 474 417 735, width=4.5 cm,height=4.4cm, clip]{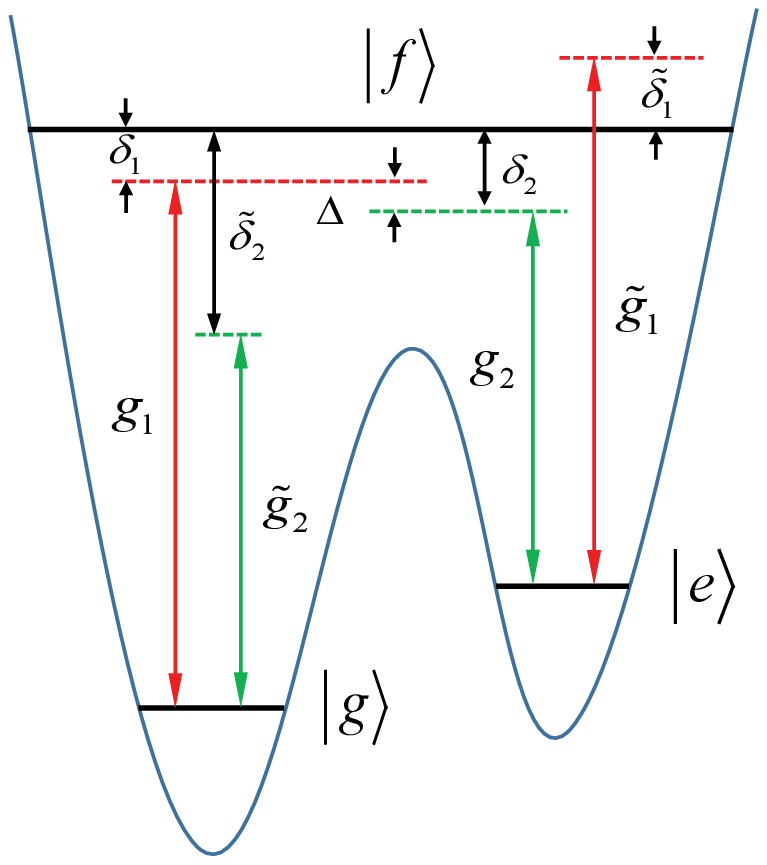}
\vspace*{-0.18in}
\end{center}
\caption{(Color online) Illustration of the unwanted coupling between cavity
$1$ and the $|e\rangle \leftrightarrow |f\rangle $ transition of the qutrit
(with coupling strength $\widetilde{g}_{1}$ and detuning $\widetilde{\protect%
\delta }_{1}$) as well as the unwanted coupling between cavity $2$ and the $%
|g\rangle \leftrightarrow |f\rangle $ transition of the qutrit (with
coupling strength $\widetilde{g}_{2}$ and detuning $\widetilde{\protect%
\delta }_{2}$). Note that the coupling of each cavity with the $|g\rangle
\leftrightarrow |e\rangle $ transition of the qutrit is negligible because
of the weak $|g\rangle \leftrightarrow |e\rangle $ transition.}
\label{fig:4}
\end{figure}

We now discuss the experimental feasibility of realizing the gate. In
reality, there exist the unwanted coupling of cavity $1$ with the $|e\rangle
\leftrightarrow |f\rangle $ transition and the unwanted coupling of cavity $%
2 $ with the $|g\rangle \leftrightarrow |f\rangle $ transition of the qutrit
(Fig.~3). After considering these factors, the Hamiltonian (2) is modified
as
\begin{equation}
\widetilde{H}_{\mathrm{I}}=H_{\mathrm{I}}+\delta H,
\end{equation}%
with%
\begin{eqnarray}
\delta H &=&\widetilde{g}_{1}(e^{-i\widetilde{\delta }_{1}t}\hat{a}%
_{1}^{+}\sigma _{fe}^{-}+h.c.)  \nonumber \\
&&+\widetilde{g}_{2}(e^{-i\widetilde{\delta }_{2}t}\hat{a}_{2}^{+}\sigma
_{fg}^{-}+h.c.).
\end{eqnarray}%
Here, $H_{\mathrm{I}}$ is the Hamiltonian (2); $\delta H$ is the
Hamiltonian, which describes the unwanted coupling between cavity $1$ and
the $|e\rangle \leftrightarrow |f\rangle $ transition with coupling strength
$\widetilde{g}_{1}$ and detuning $\widetilde{\delta }_{1}=\omega
_{fe}-\omega _{c_{1}},$ as well as the unwanted coupling between cavity $2$
and the $|g\rangle \leftrightarrow |f\rangle $ transition with coupling
strength $\widetilde{g}_{2}$ and detuning $\widetilde{\delta }_{2}=\omega
_{fg}-\omega _{c_{2}}$ (Fig.~3).

When the dissipation and dephasing are included, the dynamics of the lossy
system is determined by
\begin{align}
\frac{d\rho }{dt}=& -i[\widetilde{H}_{\mathrm{I}},\rho
]+\sum_{l=1}^{2}\kappa _{l}\mathcal{L}[a_{l}]  \nonumber \\
& +\gamma _{eg}\mathcal{L}[\sigma _{eg}^{-}]+\gamma _{fe}\mathcal{L}[\sigma
_{fe}^{-}]+\gamma _{fg}\mathcal{L}[\sigma _{fg}^{-}]  \nonumber \\
& +\sum\limits_{j=e,f}\{\gamma _{\varphi j}(\sigma _{jj}\rho \sigma
_{jj}-\sigma _{jj}\rho /2-\rho \sigma _{jj}/2)\},
\end{align}%
where $\widetilde{H}_{\mathrm{I}}$ is the above full Hamiltonian, $\sigma
_{jj}=|j\rangle \langle j|(j=e,f)$, and $\mathcal{L}[\xi ]=\xi \rho \xi
^{\dag }-\xi ^{\dag }\xi \rho /2-\rho \xi ^{\dag }\xi /2$, with $\xi
=a_{l},\sigma _{eg}^{-},\sigma _{fe}^{-},\sigma _{fg}^{-}$. In addition, $%
\kappa _{l}$ is the photon decay rate of cavity $l$ $(l=1,2),$ $\gamma _{eg}$
is the energy relaxation rate for the level $|e\rangle $ of the qutrit, $%
\gamma _{fe}(\gamma _{fg})$ is the energy relaxation rate of the level $%
|f\rangle $ of the qutrit for the decay path $|f\rangle \longrightarrow
|e\rangle (|g\rangle )$, and $\gamma _{\varphi j}$ is the dephasing rate of
the level $|j\rangle (j=e,f)$ of the qutrit. \newline

For simplicity, we consider the two qubits are initially in the following
state%
\begin{eqnarray}
|\psi _{\mathrm{in}}\rangle &=&\cos \alpha \cos \beta |0\rangle
_{1}|cat\rangle _{2}+\cos \alpha \sin \beta |0\rangle _{1}|\overline{cat}%
\rangle _{2}  \nonumber \\
&&+\sin \alpha \cos \beta |1\rangle _{1}|cat\rangle _{2}+\sin \alpha \sin
\beta |1\rangle _{1}|\overline{cat}\rangle _{2}.
\end{eqnarray}%
Thus, the ideal output state of the whole system is
\begin{eqnarray}
|\psi _{\mathrm{id}}\rangle  &=&\left( \cos \alpha \cos \beta |0\rangle
_{1}|cat\rangle _{2}+\cos \alpha \sin \beta |0\rangle _{1}|\overline{cat}%
\rangle _{2}\right.   \notag \\
&&\left. +\sin \alpha \cos \beta |1\rangle _{1}|cat\rangle _{2}-\sin \alpha
\sin \beta |1\rangle _{1}|\overline{cat}\rangle _{2}\right) \otimes
\left\vert g\right\rangle .  \notag \\
&&
\end{eqnarray}

The fidelity of the operation is defined as
\begin{equation}
\mathcal{F}=\frac{1}{\left( 2\pi \right) ^{2}}\int_{0}^{2\pi }\int_{0}^{2\pi
}\sqrt{\langle \psi _{\mathrm{id}}|\rho |\psi _{\mathrm{id}}\rangle }d\alpha
d\beta ,
\end{equation}%
where $|\psi _{\mathrm{id}}\rangle $ is the output state of an ideal system
given above, without dissipation and dephasing; while $\rho $ is the final
practical density operator of the system when the operation is performed in
a realistic situation.

\begin{figure}[tbp]
\begin{center}
\includegraphics[width=12 cm,trim=300 30 0 0,clip]{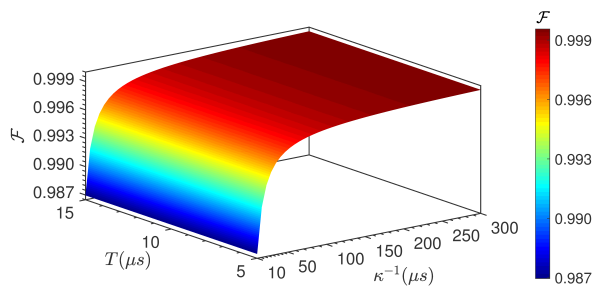} \vspace*{-0.32in%
}
\end{center}
\caption{(Color online) Fidelity versus $T$ and $\protect\kappa ^{-1}$. The
parameters used in the numerical simulation are referred to the text.}
\label{fig:4}
\end{figure}

For a flux qutrit, the typical transition frequency between neighboring
levels can be made as 1 to 20 GHz. As an example, consider $\omega
_{eg}/2\pi =5.0$ GHz, $\omega _{fe}/2\pi =7.5$ GHz, and $\omega _{fg}/2\pi
=12.5$ GHz. With a choice of $\delta _{1}/2\pi =1.5$ GHz and $\delta
_{2}/2\pi =1.65$ GHz, we have $\Delta /2\pi =150$ MHz, $\omega _{c_{1}}/2\pi
=11$ GHz, and $\omega _{c_{2}}/2\pi =5.85$ GHz. For the transition
frequencies of the qutrit and the frequencies of the cavities given here, we
have $\widetilde{\delta }_{1}/2\pi =-3.5$ GHz and $\widetilde{\delta }%
_{2}/2\pi =6.65$ GHz. Additional parameters used in the numerical simulation
are: (i) $\gamma _{eg}^{-1}=5T$ $\mu $s, $\gamma _{fe}^{-1}=2T$ $\mu $s, $%
\gamma _{fg}^{-1}=T$ $\mu $s, (ii) $\gamma _{\phi e}^{-1}=\gamma _{\phi
f}^{-1}=T$ $\mu $s, (iii) $\kappa _{1}=\kappa _{2}=\kappa ,$ (iv) $%
g_{1}/2\pi =150$ MHz, and (v) $\alpha =0.5$. According to Eq.~(11), a simple
calculation gives $g_{2}/2\pi \sim 149.8$ MHz. For a flux qutrit, $%
\widetilde{g}_{1}\sim g_{1}$ and $\widetilde{g}_{2}\sim g_{2}$. The coupling
constants chosen here are readily available because a coupling constant $%
\sim 2\pi \times 636$ MHz was reported for a flux device coupled to a
microwave cavity [25].

By solving the master equation (14), we numerically plot Fig. 4, which
illustrates the fidelity versus $T$ and $\kappa ^{-1}$. From Fig. 4, one can
see that when $T\geqslant 5$ $\mu $s and $\kappa ^{-1}\geqslant 136$ $\mu $%
s, fidelity exceeds $99.9\%$, which implies that a high fidelity can be
obtained for the gate being performed in a realistic situation.

With the parameters chosen above, the gate operational time is estimated as $%
\sim 0.37$ $\mu $s, much shorter than the decoherence times of the qutrit ($5
$ $\mu $s $-$ $75$ $\mu $s) and the cavity decay times ($10$ $\mu $s $-$ $300
$ $\mu $s) considered in Fig. 4. In our numerical simulation, we consider a
rather conservative case for decoherence time of the flux qutrit, because
experiments have reported decoherence time 70 $\mu $s to 1 ms for a
superconducting flux device [26,27]. For $\kappa ^{-1}=136$ $\mu $s and the
cavity frequencies given above, a simple calculation gives $Q_{1}\sim
9.39\times 10^{6}$ for cavity $1$ and $Q_{2}\sim 4.99\times 10^{6}$ for
cavity $2.$ Note that a high quality factor $Q=3.5\times 10^{7}$ of a 3D
microwave cavity has been experimentally demonstrated [18,28]. Our analysis
here implies that the high-fidelity implementation of the proposed gate is
feasible within the current circuit QED technology.

\section*{Funding Information}

This work was supported in part by the NKRDP of China (Grant No.
2016YFA0301802) and the National Natural Science Foundation of China under
Grant Nos. [11074062, 11374083,11774076].


\begin{thebibliography}{99}
\bibitem{s1} O. P. de S\'{a}Neto, and M. C. de Oliveira, J. Phys. B At. Mol.
Opt. Phys. \textbf{45}, 185505 (2012).

\bibitem{s2} G. Y. Wang, Q. Liu, H. R. Wei, T. Li, Q. Ai, and F. G. Deng,
Sci. Rep. \textbf{6}, 24183 (2016).

\bibitem{s3} D. Yu, M. M. Valado, C. Hufnagel, L. C. Kwek, L. Amico, and R.
Dumke, Phys. Rev. A \textbf{93}, 042329 (2016).

\bibitem{s4} S. Hoffman, C. Schrade, J. Klinovaja, and D. Loss, Phys. Rev. B
\textbf{94}, 045316 (2016).

\bibitem{s5} D. Kim and K. Moon, arxiv: 1808.02865.

\bibitem{s6} A. Reiserer, N. Kalb, G. Rempe, and S. Ritter, Nature (London)
\textbf{508}, 237 (2014)

\bibitem{s7} C. J. Ballance, V. M. Sch\"{a}fer, J. P. Home, D. J. Szwer, S.
C. Webster, D. T. C. Allcock, N. M. Linke, T. P. Harty, D. P. L. Aude Craik,
D. N. Stacey, A. M. Steane and D. M. Lucas, Nature (London) \textbf{528},
384 (2015)

\bibitem{s8} Z. L. Xiang, S. Ashhab, J. Q. You, and F. Nori, Rev. Mod. Phys.
\textbf{85}, 623 (2013).

\bibitem{s9} J. Q. You and F. Nori, Nature (London) \textbf{474}, 589 (2011).

\bibitem{s10} N. Ofek, A. Petrenko, R. Heeres, P. Reinhold, Z. Leghtas, B.
Vlastakis, Y. Liu, L. Frunzio, S. M. Girvin, L. Jiang, M. Mirrahimi, M. H.
Devoret, and R. J. Schoelkopf, Nature (London) \textbf{536}, 441 (2016).

\bibitem{s11} Y. X. Liu, L. F. Wei, and F. Nori, Europhys. Lett. \textbf{67}%
, 941 (2004).

\bibitem{s12} M. Hua, M. J. Tao, and F. G. Deng, Phys. Rev. A \textbf{90},
18824 (2014).

\bibitem{s13} A. N. Korotkov, Phys. Rev. B \textbf{84}, 014510 (2011).

\bibitem{s14} C. P. Yang, Q. P. Su, S. B. Zheng, and S. Han, Phys. Rev. A
\textbf{87}, 022320 (2013).

\bibitem{s15} M. Mirrahimi, Z. Leghtas, V. V. Albert, S. Touzard, R. J.
Schoelkopf, L.Jiang, and M.H.Devoret, New J. Phys. \textbf{16}, 045014
(2014).

\bibitem{s16} S. E. Nigg, Phys. Rev. A \textbf{89}, 022340 (2014).

\bibitem{s17} C. P. Yang, Q. P. Su, S. B. Zheng, F. Nori, and S. Han, Phys.
Rev. A \textbf{95}, 052341 (2017).

\bibitem{s18} C. Wang, Y. Y. Gao, P. Reinhold, R. W. Heeres, N. Ofek, K.
Chou, C. Axline, M. Reagor, J. Blumoff, K. M. Sliwa, L. Frunzio, S. M.
Girvin, L. Jiang, M. Mirrahimi, M. H. Devoret, and R. J. Schoelkopf, Science
\textbf{352}, 1087 (2016).

\bibitem{s19} M. Nielsen and I. Chuang, Quantum Computation and Quantum
Information (Cambridge Univ. Press, Cambridge, 2000).

\bibitem{s20} O. Morin, K. Huang, J. Liu, H. L. Jeannic, C. Fabre, and J.
Laurat, Nat. Photonics 8, 570 (2014).

\bibitem{s21} H. Jeong, A. Zavatta, M. Kang, S. W. Lee, L. S. Costanzo, S.
Grandi, T. C. Ralph, and M. Bellini, Nat. Photonics 8, 564 (2014).

\bibitem{s22} P. J. Leek, S. Filipp, P. Maurer, M. Baur, R. Bianchetti, J.
M. Fink, M. G\"{o}ppl, L. Steffen, and A. Wallraff, Phys. Rev. B \textbf{79}%
, 180511 (2009).

\bibitem{s23} M. Sandberg, C. M. Wilson, F. Persson, T. Bauch, G. Johansson,
V. Shumeiko, T. Duty, and P. Delsing, Appl. Phys. Lett. \textbf{92}, 203501
(2008).

\bibitem{s24} D. F. James and J. Jerke, Can. J. Phys. \textbf{85}, 625
(2007).

\bibitem{s25} T. Niemczyk, F. Deppe, H. Huebl, E. P. Menzel, F. Hocke, M. J.
Schwarz, J. J. Garcia Ripoll, D. Zueco, T. H\"{u}mmer, E. Solano, A. Marx,
and R. Gross, Nat. Phys. \textbf{6}, 772 (2010).

\bibitem{s26} F. Yan, S. Gustavsson, A. Kamal, J. Birenbaum, A. P. Sears, D.
Hover, T. J. Gudmundsen, J. L. Yoder, T. P. Orlando, J. Clarke, A. J.
Kerman, and W. D. Oliver, Nat. Commun. \textbf{7}, 12964 (2016). 

\bibitem{s27} J. Q. You, X. Hu, S. Ashhab, and F. Nori, Phys. Rev. B \textbf{75}, 140515(R) (2007).

\bibitem{s28} M. Reagor, W. Pfaff, C. Axline, R. W. Heeres, N. Ofek, K.
Sliwa, E. Holland, C. Wang, J. Blumoff, K. Chou, M. J. Hatridge, L. Frunzio,
M. H. Devoret, L. Jiang, and R. J. Schoelkopf, Phys. Rev. B \textbf{94},
014506 (2016).
\end{thebibliography}
\end{document}